\documentclass[final,12pt,authoryear,times]{elsarticle}
\usepackage[utf8]{inputenc}
\usepackage{amsmath}
\usepackage{mathrsfs}
\usepackage{graphicx}
\usepackage{epstopdf}
\usepackage{enumitem}
\usepackage{siunitx}
\usepackage{float}
\usepackage{mathtools}
\usepackage{xcolor}
\usepackage{subcaption}
\usepackage{algorithm}
\usepackage{algorithmic}
\usepackage[hidelinks]{hyperref}
\usepackage{textcomp} 
\newcommand{\comment}[1]{} 

 \usepackage[modulo]{lineno}

\usepackage{amssymb}
\usepackage{graphicx}
\usepackage{float}
\usepackage[english]{babel}
\usepackage[normalem]{ulem}
\usepackage{amsmath}
\usepackage{amsthm}
\usepackage{amsfonts}
\usepackage{enumerate}
\usepackage{bm}

\usepackage{physics}
\usepackage{lineno}

\journal{Journal of the Mechanics and Physics of Solids}

\begin{document}



\begin{frontmatter}

\title{Mechanical Forces Quench Frontal Polymerization: Experiments and Theory}



\author[label1]{Xuanhe Li}
\author[label1,label2]{Tal Cohen\corref{cor1}}
\cortext[cor1]{Corresponding author: talco@mit.edu}

\address[label1]{Massachusetts Institute of Technology, Department of Mechanical Engineering, Cambridge, MA, 02139, USA}
\address[label2]{Massachusetts Institute of Technology, Department of Civil and Environmental Engineering, Cambridge, MA, 02139, USA}


\begin{abstract}
Frontal polymerization is a promising energy-saving  method for rapid fabrication of polymer components with good mechanical properties. In these systems, a small energy input is sufficient to convert monomers, from a liquid or soft solid state, into a stiff polymer component. Once the reaction is initiated, it propagates as a self-sustaining front that is driven by the heat released from the  reaction itself. While several studies have been proposed to capture the coupling between thermodynamics and extreme chemical kinetics in these systems, and can explain experimentally observed thermo-chemical instabilities, only few have considered the potential influence of mechanical forces that develop in these systems during fabrication. Nonetheless, some experiments do indicate that local volume changes induced by the competing effects of thermal  expansion and chemical shrinkage, can lead to significant deformation or even failure in the resulting component.
In this work, we present a unique experimental approach to elucidate the effect of mechanics on the propagation. Our experiments reveal that  residual stresses that arise  in frontal polymerization are  not only a potential cause of undesired deformations in polymer products, but can also  quench the reaction front. This thermo-chemo-mechanically coupled effect is captured by our theoretical model, which explains the mechanical limitations on frontal polymerization and can guide future fabrication. Overall, the findings of this work suggest that  mechanical coupling needs to be taken into consideration to enable  industrial applications of frontal polymerization at large scales.
\end{abstract}


\end{frontmatter}


\section{Introduction}

Polymers and polymer-based composites have a broad range of applications in the manufacturing industry. Polymer components are fabricated through the polymerization reaction, where  monomer molecules are bonded to form chains that are cross-linked to form a network. Continuous energy input, such as heat  and pressure (thermal curing, \cite{hay2006recent}) or light (photo-polymerization,  \cite{corrigan2019seeing}), is required throughout the process to sustain and control the polymerization reaction. While autoclaves are commonly used for thermal curing, their cost increases exponentially with the size of components \citep{abliz2013curing}, which remains a great challenge in light of the increasing demand for fabrication of large-scale polymer composites in applications such as wind turbine blades \citep{mishnaevsky2017materials}, automotive components \citep{friedrich2013manufacturing}, and aircraft wings \citep{Boeing}.

In lieu of a continuous supply of external energy, Frontal Polymerization (FP)  exploits the heat released from the polymerization reaction to activate further reaction in the neighboring regions of the material. This forms  a self-sustained polymerization front which  propagates  through the  component \citep{suslick2023frontal}. As a result, FP shows great potential as a rapid, economically-efficient and scalable curing technology \citep{robertson2018rapid}. Frontal polymerization processes can be initiated from either liquid state or partially cured gel-like solid state, which further broadens its application to 3D printing \citep{aw2022self}, surface patterning \citep{kumar2022surface} and mold-free fabrication through deformation and embossing \citep{robertson2018rapid}.

Polymerization processes are typically accompanied by volume change that is induced  by competing effects of thermal expansion and chemical shrinkage.  In various applications, such as polymer coatings \citep{francis2002development}, composite laminates \citep{bogetti1992process}, and textile polymer composites \citep{heinrich2013role}, the polymerization process can thus lead to volume mismatch between the resin and the substrate or the embedded fibers, resulting in residual stresses which can induce failure, such as delamination and cracks, and can significantly reduce the geometrical precision and performance of components. 

While several theoretical models and experimental investigations   have been devoted to understand and predict mechanical effects  in bulk  curing processes \citep{sain2018thermo, wu2018evolution,wang2023shrinkage}, less is known about this effect in FP.  In the latter,  a high gradient zone of the polymerization front separates the cured and the uncured regions of the sample.  Both the temperature and the degree of curing  can vary rapidly across the front and thus a localized zone of significant volume mismatch will form and will introduce localized stresses that can lead to deformation or even cracks in the component \citep{binici2006spherically}. To understand and mitigate these effects in FP, it is essential to extend existing  models  (such as the one by \cite{goli2018frontal}) to include also the coupled role of mechanics. 

In addition to volume, another mechanical property that changes significantly across  
the polymerization front  is the stiffness. With new chains formed and cross-linked to the polymer network, the material stiffness can increase by four orders of magnitude during the process. As indicated in earlier studies  \citep{gillen1988effect}, an important feature of the newly formed cross-links is that they can be assumed to form in a stress-free state. As a result, if deformation is introduced during the curing process, the macroscopic stress-free configuration will also evolve. To capture this effect in bulk polymerization, \cite{hossain2009small}  developed a hypoelastic representation of the  constitutive response whereby the relation between stress and strain is defined  incrementally in  a rate form. In this paper, we will develop a new method that captures  the evolution of the stress-free state in FP by introducing a transformation strain into the kinematic description and prescribing its evolution using a thermodynamically consistent kinetic law. 

When conducting FP in a sample that is initially a soft  gel, the  process can be considered as a solid-solid phase transformation with a moving phase boundary, which is similarity to various phase transformation systems with examples ranging from martensitic transformations \citep{abeyaratne1993continuum} to biological growth \citep{abi2019kinetics}. To model such phase transformation systems at the continuum level, a commonly used approach is to introduce a discontinuous phase boundary whose motion is determined by a kinetic law \citep{abeyaratne2006evolution}, which relates the velocity of the phase boundary to its thermodynamic conjugate - a  driving force, while obeying the  second law of thermodynamics. As a result, the motion of the phase boundary may be influenced directly by the local stress-state, as observed in various phase change systems, including the confined growth of bacterial biofilm  and tumors  \citep{li2022nonlinear, senthilnathan2023large} and   the diffusion of Li-ion in electrode materials \citep{di2014cahn}. For FP processes, however, it  remains unclear whether the stress-state  influences the propagation behavior of the polymerization front. In a recent study by \cite{kumar2022surface},   an indirect (one-way) coupling between mechanics and the reaction process  was considered, such that  simultaneous deformations of the sample, during its curing, influence the eventual form of the component, but the potential two-way coupling effect was not considered.

The main objective of this paper is to investigate the effect of the stress-state on the propagation dynamics of the polymerization front during FP process. We achieve this by a combination of  experimental and theoretical tools. In the next section (Section 2), we thus  proceed to desribe  our experimental setup and our  approach to isolate the mechanical influence on the propagation. We then present our experimental results which reveal that mechanical coupling indeed has a direct influence on the propagation. Then, in Section 3, we develop a fully coupled theoretical framework considering uniaxial motion. In Section 4 we present results obtained through numerical integration of this model, which requires a moving mesh method to capture the rapid change of state variables
within the high gradient zone of the front. Results are shown to explain our experimental observations of the coupled  phenomenon. 
Finally, in Section 5 we provide some concluding remarks.





\section{Experiments}
The key to  experimental confirmation of the mechanical influence on the front propagation in FP is in isolating the mechanical effect from other factors, such as 
heat loss. Here we describe the  material system that we used and the experimental procedure that we developed to achieve this goal.

\subsection{Sample preparation}
\begin{figure}[H]
    \centering
    \includegraphics[width=1\linewidth]{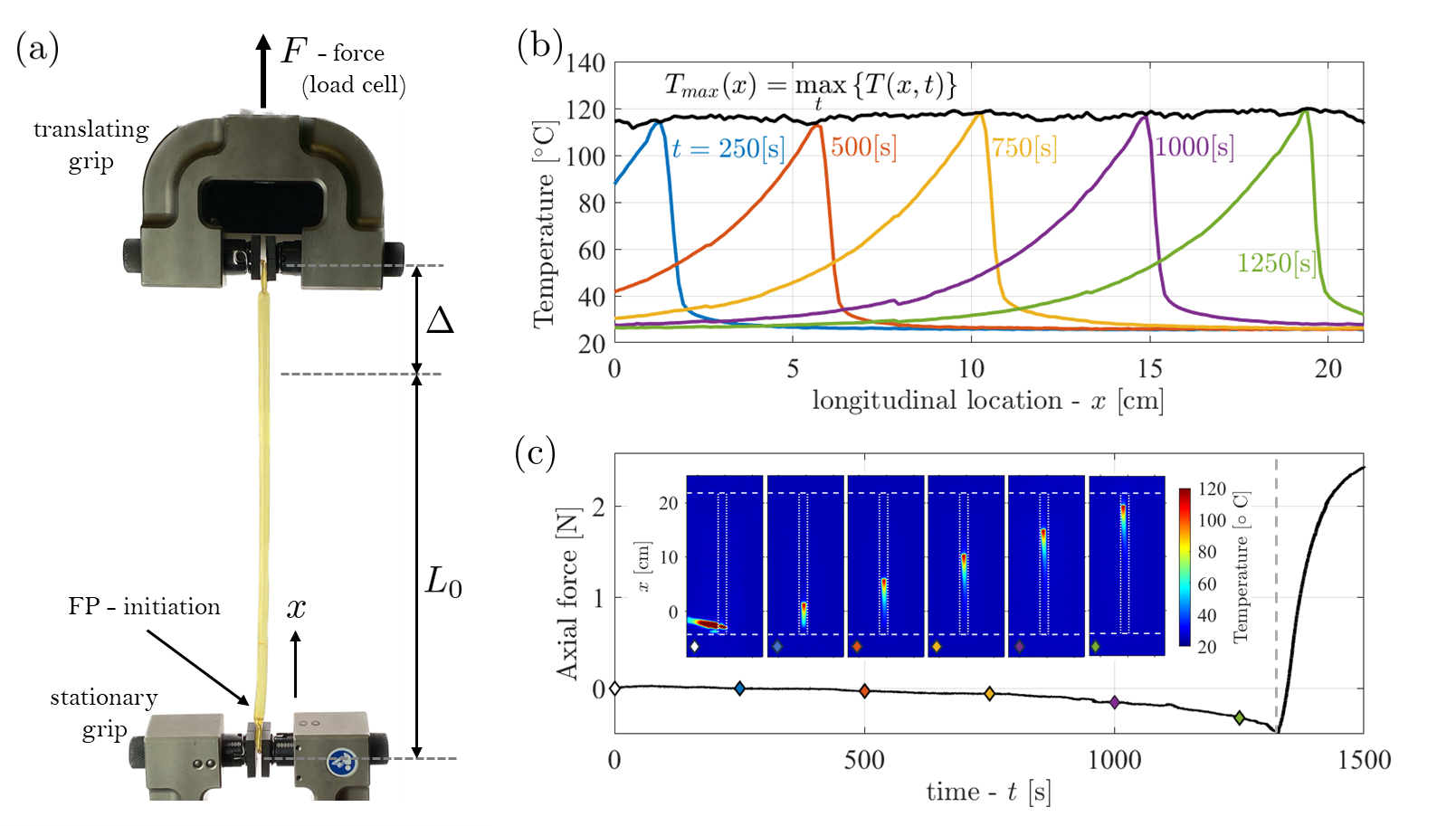}
    \caption{(a) Experimental setup - cylindrical DCPD gel sample mounted on Instron machine. The diameter of the cylindrical sample shown in this image is $8$ mm; b) Temperature profile along the central axis of the sample at different times for the case without pres-stretch $(\bar{\varepsilon}=0)$. Black curve corresponds to the maximum temperature experienced at a given $x$ location throughout all times, as indicated by the formula; c) Curve shows axial force $F$ as a function of time (grey dashed line indicates the time when the sample was fully polymerized). Insets show a sequence of infrared images taken at times as indicated by colored markers on the axial force curve and corresponding to the temperature distributions in (b).  A video showing the entire propagation is provided in the Supplementary Information (\href{https://www.dropbox.com/scl/fi/pakdtpsqn99rg1ayhwomy/FP_Fig_1_propogation_stretch_free.avi?rlkey=5v8hhdywpkffubiwomar75xmp&dl=0}{\textcolor{blue}{see link}}).}
    \label{fig:experiment}
\end{figure}

In this work  we fabricate Dicyclopentadiene (DCPD) gels following the  material preparation method  described in \cite{robertson2018rapid}. As detailed therein, DCPD (contains BHT as stabilizer), Grubbs catalyst $\#$M204 (GC2), Cyclohexylbenzene $97\%$, Tributyl phosphite $93\%$(TBP) and 5-Ethylidene-2-norbornene $99\%$(ENB) were purchased from Sigma-Aldrich. 
For a typical test, 13.5 $\mathrm{mg}$ GC2 catalyst powder was dissolved with $1.68\mathrm{mL}$ cyclohexylbenzene, and then mixed with $8\mathrm{mg}$ inhibitor TBP. Then $20\mathrm{g}$ DCPD (solid at room temeprature) was  melted on a hotplate at $40^\circ $C and  mixed with $1.05 \mathrm{g}$ ENB to obtain a 95:5 DCPD:ENB mixture, which is liquid at room temperature. Both the catalyst/inhibitor mixture and the DCPD:ENB mixture were  degassed for 30 minutes before being mixed together thoroughly. The liquid mixture was also degassed for an additional 30 minutes before being poured into several thin glass tubes of varying diameters {($ 8 - 12$ mm)}  to create slender cylindrical samples with lengths in the range  {($10 - 30$ cm)}.  The filled tubes are then left to rest at room temperature to  partially cure for 18 hours to form a soft gel\footnote{Note that specialized tubes were used to enable horizontal placement, thus eliminating emergence of variations along the height of the sample.}, after which the specimen is carefully removed from the glass mold. A typical  specimen is shown in Fig.\ref{fig:experiment}(a). {The Young's modulus of the resulting soft gel samples, prior to FP, was $\sim 200$ kPa, as estimated from the experiments (see \ref{gel}).   }

\subsection{Experimental setup and FP procedure}
The soft gel-like solid specimens were mounted on an Instron universal testing machine,  as shown in Fig. \ref{fig:experiment}(a). The slender geometry of the specimen is chosen to enable a  simplified uniaxial representation of the field. The Instron allows us to control the displacement of the sample by vertical translation of the top grip, while the bottom grip remains stationary. Simultaneously, the applied force, $F(t)$, is measured via a load cell throughout the process, where $t$ denotes time. We denote the initial length of the specimen by $L_0$ and the displacement of the top grip is denoted by $\Delta$, so that the macroscopic applied strain in the sample is  $\bar{\varepsilon} = \Delta/L_0$, which must be distinguished from the local strain, $\varepsilon(x)$, that can change throughout the FP process along the sample as denoted by the coordinate $x$ (as will be explained in the next sections). 
With the two ends of the sample gripped, the frontal polymerization process was  initiated by contacting the bottom end of the sample $(x=0)$ with a soldering iron ($\sim 300^\circ$C). The polymerization  front then propagates. An infrared camera (Omega TI-125, $320\times 240 $ pixels) was used to monitor the temperature distribution  with a frame rate of 1 frame/sec. 

\subsection{Experimental results}
As a first step, we examine the FP process in absence of applied strain, i.e. with $\bar{\varepsilon}=0$, as shown in Fig. \ref{fig:experiment}(b,c) for a sample with cross section radius $4\mathrm{mm}$ and initial length $L_0=20\mathrm{cm}$. Evolution of the temperature distribution captured by the infrared camera is shown in the inset of Fig. \ref{fig:experiment}(c) and provided in the Supplementary Information (\href{https://www.dropbox.com/scl/fi/pakdtpsqn99rg1ayhwomy/FP_Fig_1_propogation_stretch_free.avi?rlkey=5v8hhdywpkffubiwomar75xmp&dl=0}{\textcolor{blue}{see link}}). Corresponding temperature profiles taken  along the central axis are presented in Fig. \ref{fig:experiment}(b)   where a travelling wave propagation behavior is apparent. Here the maximum temperature $T_{max}$, which serves as an  envelope of all temperature profiles, is also plotted (black curve) and  shows that the propagation arrives at a steady state with a consistent peak temperature. Note that at the trailing end of the propagating wave, the temperature decreases back to room temperature, which is due to  heat loss to the environment. 

\begin{figure}[H]
    \centering
    \includegraphics[width=1\linewidth]{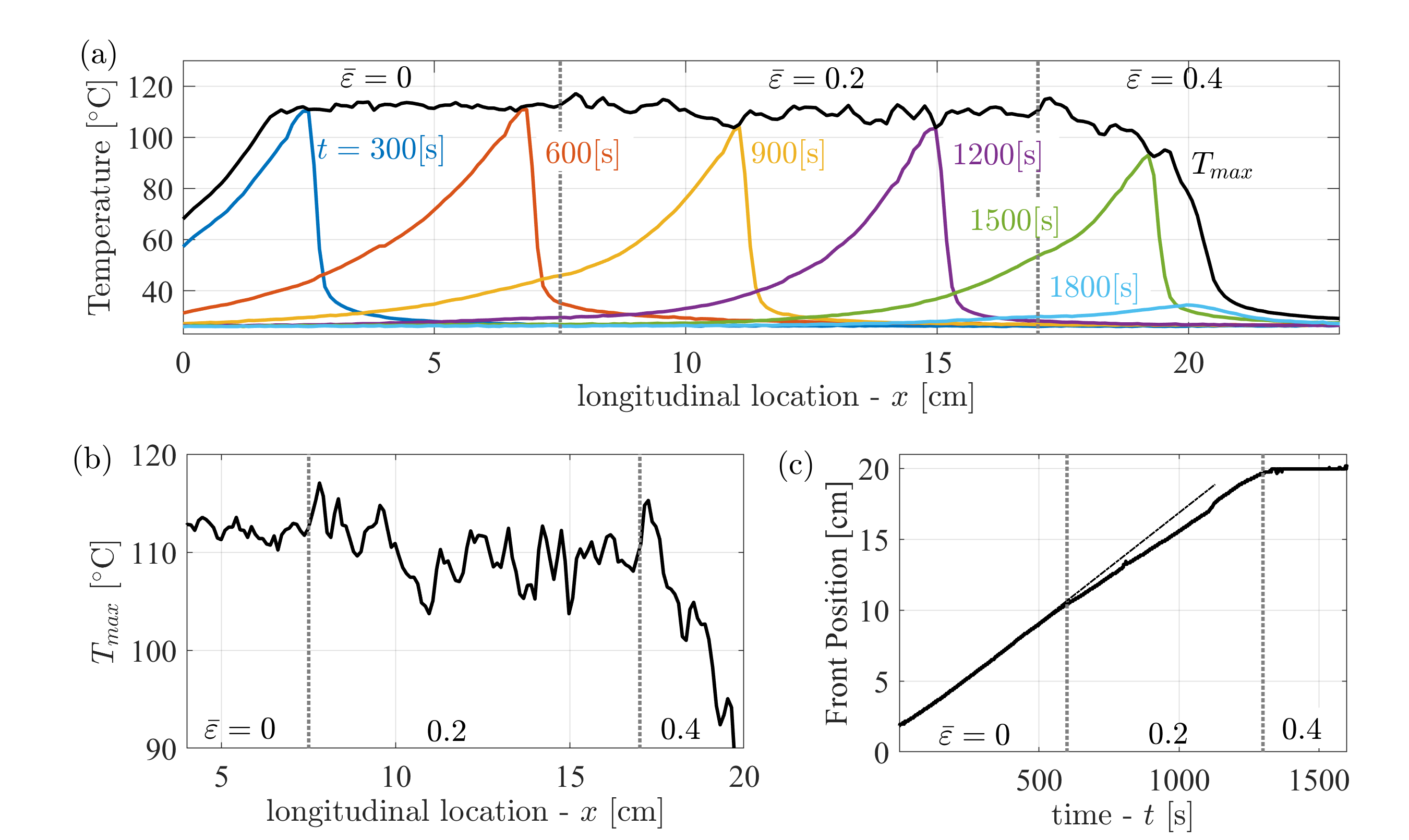}
    \caption{Experimental results for frontal polymerization with simultaneous application of strain. Strain is applied in three intervals. Initiation occurs in a stress-free sample $(\bar{\varepsilon}=0)$, then the sample is subjected to $\bar{\varepsilon}=0.2$ and later to $\bar{\varepsilon}=0.4$, as marked on the figures via dotted lines: (a) Temperature profile along the central axis of the sample at different times with $T_{max}$ corresponding to the maximum temperature at a given location, for all times; (b) Expanded view of the maximum temperature $T_{max}$ distribution; (c) Front position as a function of time. The dashed-dotted line shows the change of slope and thus indicates a change of propagation velocity upon stretching. A video showing the entire propagation is provided in the Supplementary Information (\href{https://www.dropbox.com/scl/fi/dpye593ursunknxbpr2vt/FP_Fig_2_propogation_stretch.avi?rlkey=kbi5rtl5dhlpdep31isajwke1&dl=0}{\textcolor{blue}{see link}}).}
    \label{fig:STRETCH}
\end{figure}

In Fig. \ref{fig:experiment}(c) we also present the axial force change $F(t)$ measured by the Instron. As the front propagates, heating and thermal expansion induce a compressive  force. A rapid transition is observed at the moment when the sample has been fully polymerized (grey dashed line), and eventually  the axial force becomes tensile. While this transition can be partially attributed to the temperature decrease due to heat dissipation, the fact that the force remains positive indicates that residual stress has been accumulated as front propagates. This persistence of positive force values cannot be exclusively accounted for by the thermal expansion, underscoring the presence of additional factors contributing to the observed mechanical behavior.


Next, to investigate the influence of applied strain on the propagation, we examine the FP process in stretched samples. In Fig. \ref{fig:STRETCH}(a) we show experimental results analogous to those in  Fig. \ref{fig:experiment}(b), but where the sample was subjected to varying levels of applied strain throughout FP. Initiation occured from a stress-free state $(\bar{\varepsilon}=0)$,  then the sample was stretched to $\bar{\varepsilon}=0.2$ and $0.4$ at discrete  time intervals, as indicated in the figure by the dotted lines. The first increase in strain, is accompanied by an abrupt change in propagation velocity, as seen from the change in slope of the curve in Fig. \ref{fig:experiment}(c). Upon further loading, to $\bar{\varepsilon}=0.4$, the propagation is fully quenched and front location remains constant (see infrared video in the Supplementary Information  
 - \href{https://www.dropbox.com/scl/fo/1b0ocke6zbd6osmvkmxmm/h?rlkey=a4whkjq5rxcwmxn5ufe3gbywe&dl=0}{\textcolor{blue}{link}}). 
 The maximum temperature distribution is also affected by the loading, as seen in Fig. \ref{fig:experiment}(b). Stretching the sample to $\bar{\varepsilon}=0.2$ induces larger  fluctuations in $T_{max}$, which is later seen to decay to room temperature at $\bar{\varepsilon }= 0.4$, as the reaction front is quenched.

While this experiment reveals a significant cross-talk between mechanical deformation and frontal polymerization, it is not possible to infer from this experiment if the effect is due to a direct coupling, or a consequence of other factors. In particular, in our slender sample geometry, heat loss to the environment plays a significant role.  In our system, loading of the sample changes its cross-section radius (through the Poisson's effect). Heat loss through the lateral surface of the body is proportional to the lateral surface area per unit volume, and thus inversely proportional to the cross-section radius.  In other words, elongation of the sample (positive strain)  accelerates  heat loss and therefore may slow the propagation, thus providing a potential explanation of the results in  Fig. \ref{fig:STRETCH}(c). 

\begin{figure}[H]
    \centering
    \includegraphics[scale = 0.6]{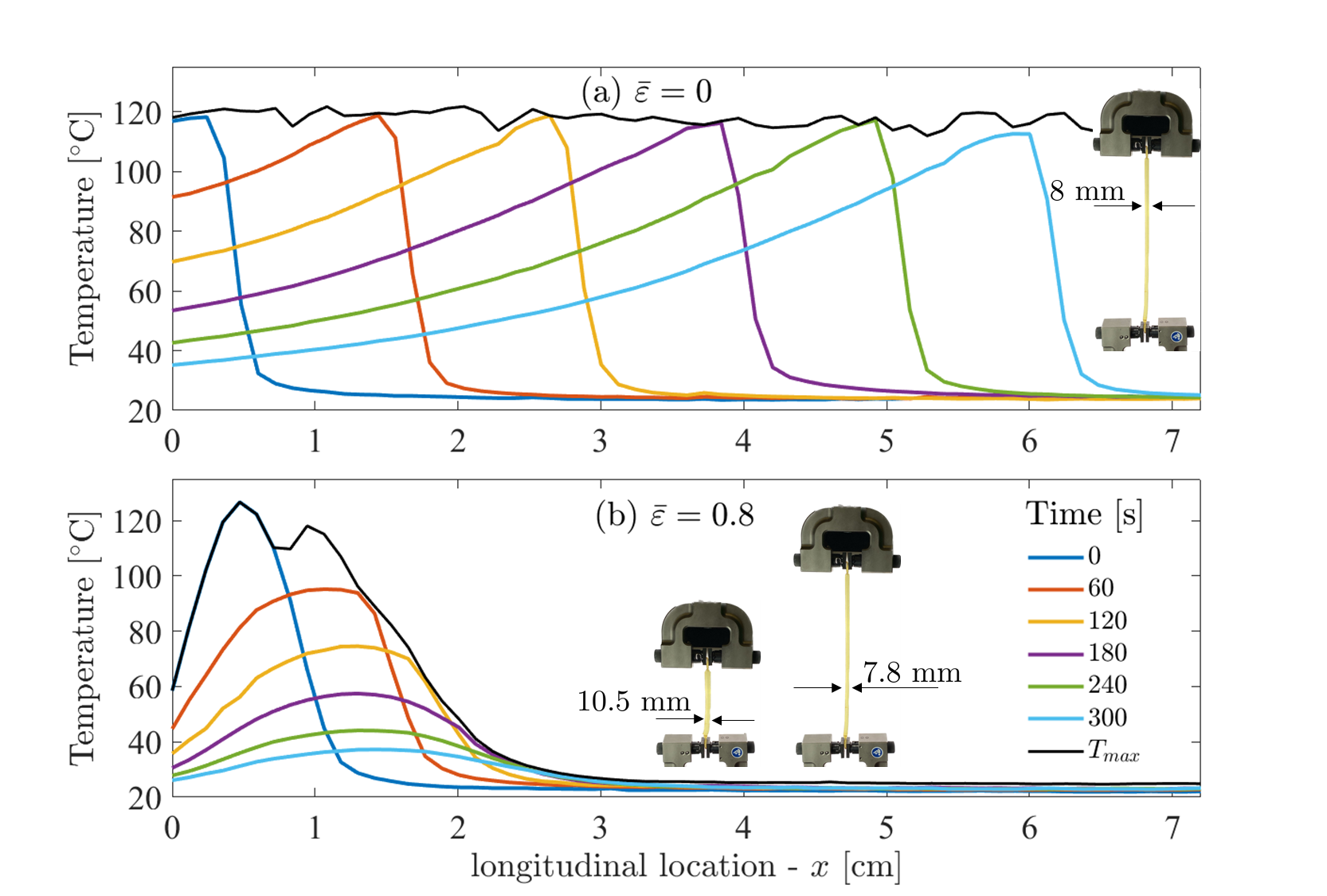}
    \caption{Experimental results for frontal polymerization with different stress-states. Two samples with cross-section diameters $8 \mathrm{mm}$ and $10.5 \mathrm{mm}$ were prepared. a) Evolution of the temperature profile with $\bar{\varepsilon}=0$ for the sample with cross-section diameter $8 \mathrm{mm}$; b) Evolution of the temperature profile with $\bar{\varepsilon}=0.8$ for the sample with initial cross-section diameter $10.5 \mathrm{mm} $ and deformed cross-section diameter $7.8 \mathrm{mm} $ after stretching. }
    \label{fig:COMP}
\end{figure}

To confirm that there is a direct influence of stress/strain on the frontal propagation, the geometric effect must be excluded. Hence,  we design a comparison experiment where two  specimens with different cross-section diameters (one is $8\mathrm{mm} $ while the other is $10.5\mathrm{mm}$) were fabricated from the same batch. The thicker sample was pre-stretched to $\bar{\varepsilon}=0.8$ while the the thinner  sample was stress-free $(\bar{\varepsilon}=0)$. The deformed diameter of the stretched sample was $7.8\mathrm{mm}$, which is approximately the same as the unstretched sample. As a result, the two samples have the same test geometry and should exhibit similar heat loss behavior, while the only difference  between the samples is the stress-state. 

The comparison of evolution of temperature profiles in the two samples,  shown in Fig. \ref{fig:COMP}, exhibits a clear influence of mechanical forces on the propagation behaviour: in the stress-free sample, the polymerization front arrives at an approximately steady state as it continues to propagate through the sample; in the pre-stretched sample, the  polymerization front is not able to sustain itself and the polymerization reaction quenches. This comparison  establishes a direct relationship between  the local stress-state and  the propagation of a polymerization front; it therefore confirms that the highly coupled interaction between mechanics, chemistry, and thermodynamics must be considered in the modeling of FP processes. In the next section we will present a theoretical model that can capture these effects. 

\section{Theory}\label{sec:theory}
In this section, we  construct a continuum model to explain the thermo-chemo-mechanically coupled phenomenon that we observe in the frontal polymerization system. The slender geometry of our specimens (as shown in Fig. \ref{fig:experiment}a) and the observed propagation of a nearly planar front,  permit a simplified uniaxial representation to  capture the main features of the process. Accordingly, all field variables can be described as functions of the material coordinate - $x$, and time - $t$. The initial cross-sectional area, $A_0$, is  defined from the pre-stretch of the sample, and the deformed area is denoted by $A$. Additionally, motivated by our observations,  we restrict our analysis  to consider small changes in strain throughout  polymerization process (i.e. after application of finite pre-stretch), and thus consider linearly elastic material response\footnote{Note that here we neglect contributions from nonlinear effects associated with large strains that are imposed as a pre-stretch to the system. }.

\subsection{Kinematics}
The motion of a particle from its initial stress-free position, $x$, to its current location, $y$, at time $t$, is described by its axial displacement - $u(x,t)$, such that $y =x+u(x,t)$.  The resulting strain  is then defined as\footnote{The subscript $x$ is used to denote partial derivatives along the material coordinate.  } $\varepsilon =u_x$. 
Other than elastic deformation - $\varepsilon_e$, 
thermal expansion - $\varepsilon_T$, can occur as a result of changes in temperature, while the curing reaction can induce contraction due to density changes as well as irreversible deformation caused by existence of stress during polymerization; both  processes contribute to the transformation strain - $\varepsilon_*$. Within the limits of linear elasticity, we can write the total strain as a decomposition of these contributions, in the form 
\begin{equation}\label{strain_decomposition}
    \varepsilon \equiv u_x = \varepsilon_e+\varepsilon_T +\varepsilon_*
\end{equation}

\subsection{Mechanical equilibrium}
In absence of transverse loads in the uniaxial setting, and neglecting the influence of body forces and inertial effects,  mechanical equilibrium simplifies to 
\begin{equation}\label{m_eq}
  F_x = 0
\end{equation}
Hence, throughout the polymerization process, the longitudinal force does not vary along $x$ coordinate and thus $F=F(t)$.

\subsection{The first law of thermodynamics - energy conservation}
The various concurrent processes occurring in our system expend energy. The chemical reaction expends internal energy (per unit length) - $U(x,t)$, to transform the monomers into polymer while also inducing heat and elastic deformation. Heat flux along the sample -  $q(x,t)$, and through its lateral surfaces - $r(x,t)$,  allows energy  to diffuse to neighboring regions along the sample and thus to further propagate the exothermic reaction and also to exchange heat with the environment. Considering a longitudinal element of unit volume, we can thus write conservation of energy in the form,  
\begin{equation}\label{EC}
    \dot{U} = F \dot{\varepsilon} - q_x -r
\end{equation}
where the superimposed dot denotes the material time derivative.

\subsection{The second law of thermodynamics - constitutive relations}
Following the Coleman-Noll methodology \citep{noll1974thermodynamics}, we  treat the second law of thermodynamics as a restriction on thermodynamically-consistent constitutive relations. 
To address the second law, we first introduce the entropy per unit length - $S(x,t)$, 
to write the Clausius–Duhem inequality as
\begin{equation}\label{CDI}
\dot S\geq -\dv{}{x}\left(\frac{q}{T}\right) -\frac{r}{T}  
\end{equation}
where $T(x,t)$ is the temperature field.
Alternatively, for the coupled system, it is convenient to introduce the Helmholtz free energy density $\psi$ (per unit length), which is the Legendre transformation of the internal energy density 
\begin{equation}\label{U_TS}
    \psi = U-TS
\end{equation}
We can then rewrite the  the Clausius–Duhem inequality \eqref{CDI}, by using \eqref{EC} and \eqref{U_TS},  in the form
\begin{equation}\label{SL}
    \dot\psi -F\dot{\varepsilon} + S\dot T+\frac{q}{T} T_x \leq 0
\end{equation}

We separate the free energy density into two contributions:  $\psi_r$  depends on the current state of the material unit,  i.e. the elastic deformation, the temperature and the degree of curing; and   $\psi_*$  accounts for the unrecoverable energy stored in the network and thus depends on the history of the process and is a direct consequence of the transformation strain  $\varepsilon_*$.
Accordingly, we assume that the Helmholtz free energy has the form
\begin{equation}
    \psi = {\psi}_r (\varepsilon_e,T,\alpha)+\psi_{*}
\end{equation}
In this work, without loss of arbitrariness, we define $\alpha$ as the normalized degree of curing, which is a linear projection of the dimensional degree of curing\footnote{Note that since our experiments are initialized from a gel state, which has already been partially cured. The initial dimensional degree of curing is thus not zero.} on the interval $[0,1]$. 

Substituting the above relation along with  \eqref{strain_decomposition} into \eqref{SL} and considering an  arbitrary dependence of the thermal strain on the temperature, i.e.  $\varepsilon_T=\varepsilon_T(T)$, we can rewrite the  Clausius–Duhem inequality,  by applying the chain rule, in the form
\begin{equation}\label{SL1}
    \left(\pdv{\psi_r}{\varepsilon_e}-F\right)\dot{\varepsilon}_e +\left(\pdv{\psi_r}{T}-F\dv{\varepsilon_T}{T}+S\right) \dot T+\frac{q}{T}T_x +\left(\dot{\psi_*}-F\dot{\varepsilon}_*\right)+\pdv{\psi_r}{\alpha}\dot\alpha \leq 0
\end{equation}
which should hold for any arbitrary set $\dot\varepsilon_e$, $\dot T$, $\dot \alpha$, $\dot{\varepsilon}_*$, and $T_x$. As a result, we require that each of the terms must separately obey the inequality. Note that each term in \eqref{SL1} represents a dissipative process of the chemo-thermo-mechanically coupled system: the first and  second terms represent  rate-dependent effects of the mechanical and thermal response of the material, respectively; the third term represents heat conduction; the fourth term corresponds to heat generated by irreversible deformations $\varepsilon_*$; and the fifth term represents the polymerization reaction.

For the present material system we neglect rate effects on the material response. Accordingly, the first and second terms in \eqref{SL1} represent conservative processes and thus vanish. Constitutive relations follow as 
\begin{equation}\label{sigma}
    F(\varepsilon_e,T,\alpha) =\pdv{\psi_r}{\varepsilon_e}
\end{equation} for the longitudinal force\footnote{Recall that the longitudinal force is a measurable quantity in our experimental system. Though at a given time it is uniform throughout the sample, its value changes over time as shown in Fig. \ref{fig:experiment}(c).}, and 
\begin{equation}\label{S_xt}
    S(\varepsilon_e,T,\alpha) = -\pdv{\psi_r}{T}+F\dv{\varepsilon_T}{T}
\end{equation} for the entropy. 
The third and second terms in \eqref{SL1} imply the inequalities
\begin{equation}\label{heat_ineq}
    q T_x\leq0 \qquad \text{and} \qquad \dot{\psi_*}\leq F{\dot\varepsilon_*}~,
\end{equation}
respectively, considering that temperature is always positive.

Finally, the last term in \eqref{SL1} implies 
\begin{equation}
 \pdv{\psi_r}{\alpha} \dot\alpha \leq 0
\end{equation}
From the above inequality, we identify a thermodynamic driving force -  $\mathcal{F}$, which is conjugate to the  rate of the chemical reaction - $\dot \alpha$  
\begin{equation}\label{driving_force}
    \mathcal{F} (\varepsilon_e,T,\alpha) = -\pdv{\psi_r}{\alpha}
\end{equation}
As a result, the second law \eqref{SL1} provides a restriction on the chemical kinetic relation  such that $\mathcal{F} \dot \alpha \geq 0$. If we define  $\dot \alpha =\mathcal{G}(\varepsilon_e,T,\alpha)$, this implies 
\begin{equation}\label{alpha_ineq}
    \mathcal{F} \cdot \mathcal{G} \geq 0
\end{equation}

\subsection{Heat equation}
To write the heat equation for the coupled system, we  first define the commonly used heat capacity $c$ in terms of entropy
\begin{equation}\label{capacity}
    c = T \frac{\partial S}{\partial T}
\end{equation}
With the constitutive relation for entropy \eqref{S_xt}, we obtain the internal energy, $U$, from \eqref{U_TS}. The energy balance equation \eqref{EC} can then be rewritten to describe the evolution of temperature in the form
\begin{equation}\label{heat0}
c\dot T =  \mathcal{F}\dot{\alpha} -T\left(\pdv{S} {\alpha}\dot{\alpha}+\pdv{S} {\varepsilon_e}\dot{\varepsilon}_e\right)   -q_x -r+\left(F\dot{\varepsilon}_*-\dot{\psi}_*\right)
\end{equation}


Note that the last term in \eqref{heat0} represents the heat generated due to the transformation strain $\varepsilon_*$, which is always larger or equal to zero according to \eqref{heat_ineq}.


\subsection{Constitutive response functions}
With the above restrictions on the constitutive relations, we now proceed to define the specific constitutive response functions. In absence of experimental data to motivate a specific form of the free energy function, here we employ the simplest mathematical representation by the decomposition
\begin{equation}\label{free_enrg}
    \psi_r(\varepsilon_e,T,\alpha) =\psi_e(\varepsilon_e,\alpha)+\psi_T(T)+\psi_{c}(\alpha)
\end{equation}
where $\psi_e$ is the elastic energy, $\psi_T$ is the thermodynamic energy, and $\psi_c$ is the chemical  energy. This partition into deformation related and temperature/chemical related  parts was considered earlier by \cite{erbts2015partitioned}.

The linear elastic energy density takes the quadratic form 
\begin{equation}\label{modulus}
    \psi_e(\varepsilon_e,\alpha) =\frac{1}{2} E(\alpha) \varepsilon_e^2 \qquad \text{with} \qquad {E}(\alpha) = E_0+E_H\alpha
\end{equation}
Here we consider also a linear dependence of the longitudinal stiffness, $E(\alpha)$, on the curing ratio, where $E_0$ is the initial longitudinal stiffness measured from the gel state
($\alpha =0$) and $E_H$ is the  chemical hardening modulus\footnote{Note that in the present uniaxial framework, these modulii are not the Young's modulus and have units of force.}.

Our experiment  shows that the longitudinal modulus, $E$, increases by 4 orders of magnitude during  the polymerization process (details in \ref{gel})  which indicates that $E_H \gg E_0$. 
Note  that although the considered polymeric material  may exhibit material nonlinearity and entropic elasticity such that the elastic modulus is influenced by temperature, we neglect these effects in comparison with the large  changes in elastic modulus due to chemical reaction. 

Next, the chemical energy related to the polymerization reaction is also assumed to take a quadratic form
\begin{equation}\label{chem_eng}
    \psi_c(\alpha) = \frac{H}{2}(1-\alpha)^2
\end{equation}
where the chemical modulus - $H$ is constant in this model, thus  neglecting direct effect of temperature or stress on the reaction heat. This form of chemical  energy introduces an energy well (i.e. a minimal value) at $\alpha =1$, which favors the polymerization process in the range  $\alpha \in [0 ,1]$. 

To describe the influence of temperature on the free energy, we depart from the definition of heat capacity $c$ given in \eqref{capacity}, and adopt the form  \citep{holzapfel1996entropy,loeffel2011chemo, mehnert2016nonlinear, mehnert2017towards} 
\begin{equation}\label{therm_eng}
    \psi_T(T) = c\left[\left(T-T_0\right)-T \mathrm{ln}\left(\frac{T}{T_0}\right)\right]
\end{equation}
where $c$ - the  heat capacity per unit length, is assumed to be a constant, and $T_0$  is the room temperature.
For the thermal strain $\varepsilon_T$ we consider a linear dependence on the temperature to write
\begin{equation}\label{eT}
    \varepsilon_T =\beta (T-T_0)
\end{equation}
where $\beta$ is the thermal expansion coefficient. Note that  the above two constitutive functions along with  \eqref{S_xt} identically satisfy the definition in \eqref{capacity} with a constant $c$.

Inserting  the specific forms of the free energy components \eqref{modulus}-\eqref{therm_eng} in \eqref{free_enrg} and using  \eqref{eT}  in the constitutive relations  \eqref{sigma}, \eqref{S_xt} and \eqref{driving_force}, reads
\begin{equation} \label{const}
\begin{aligned}
    F(\varepsilon_e,\alpha) &=(E_0 +E_H\alpha) \varepsilon_e \\
    S(\varepsilon_e,T,\alpha) &= c \ln\left(\frac{T}{T_0}\right) + \beta F(\varepsilon_e,\alpha)\\
    \mathcal{F} (\varepsilon_e,\alpha)&= H (1-\alpha) -\frac{1}{2}E_H \varepsilon_e^2 
\end{aligned}
\end{equation}
 for the applied force, the entropy, and the driving force, respectively. 

Notice that in the  stress-free state ($\varepsilon_e = 0$), the driving force $\mathcal{F}(0,\alpha) = H(1-\alpha)\geq0$ for all $\alpha\in[0,1]$. This indicates that the dissipation inequality \eqref{alpha_ineq} is always satisfied  for any irreversible chemical kinetic relation $\dot\alpha = \mathcal{G}(\varepsilon_e,T,\alpha)\geq0$. Hence, the polymerization reaction is a spontaneous process under stress-free conditions. 
However, if the elastic strain $\varepsilon_e$ is large enough,  the driving force may become  negative, in which case $\dot\alpha\equiv 0$ to ensure the process is irreversible while satisfying  the dissipation inequality. To take this restriction into account, we adopt  the methodology used by \cite{abeyaratne2006evolution}, and choose the chemical kinetic function in the form 
\begin{equation}\label{chem_kinetic}
    \mathcal{G}(\varepsilon_e,T,\alpha) = \left\{
\begin{aligned}
     Ce^{-\frac{E_a}{RT}}\left(\frac{\mathcal{F}(\varepsilon_e,\alpha)-\mathcal{F}_0(T,\alpha)}{H}\right)^n&\ &\mathcal{F}\ge \mathcal{F}_0 \\
    0\qquad\qquad&\ &\mathcal{F}< \mathcal{F}_0\\
\end{aligned}    
\right.
\end{equation}
where  $C$ is the  reaction rate constant. The temperature dependence of the reaction rate is described by the commonly used Arrhenius relation, where $E_a$ is the activation energy and $R$ is the universal gas constant.
A power law dependence with $n>0$    guarantees the continuity of the kinetic relation at $\mathcal{F}=\mathcal{F}_0$, where  $\mathcal{F}_0$ is introduced to capture  the activation barrier  to the polymerization reaction
\begin{equation}\label{barrier}
\mathcal{F}_0(T,\alpha) =   h(1-\alpha)  \frac{E_a}{RT}
\end{equation}
which accounts for the influence of the local temperature $T$, and the remaining concentration of reactant  $(1-\alpha)$, on the propensity to react, 
with  magnitude coefficient - $h$.

To describe the heat transfer within the sample, while ensuring that the inequality \eqref{heat_ineq} is satisfied, we employ the commonly used Fourier's law
\begin{equation}\label{Fourier}
    q = -\kappa T_x
\end{equation}
where $\kappa >0$ is the thermal conductivity. 

Heat loss to the environment is considered as an external heat sink in our system and is thus not subjected to a thermodynamic restriction. Our experiments (see \ref{heat_loss}) confirm that heat loss is well captured by Newton's law of cooling, which can be written as 
\begin{equation}\label{newton} r = \lambda(T-T_0),\end{equation}
where $T_0$ is the room temperature and $\lambda$ - the heat loss coefficient is an experimentally measured quantity (see \ref{heat_loss}).

\subsection{Transformation strain}
To complete the constitutive representation of our system, it remains to provide an evolution relation for  the transformation strain as a function of the state variables. It is instructive to notice the analogy between the transformation strain considered here and the plastic strain in plasticity theory, in the latter  a \textit{flow rule} serves as an evolution law and is motivated by the specific plasticity mechanism in the material.  Similarly, in this work, we motivate the choice of an evolution law by considering the specific mechanism that leads to development of the transformation strain.  

\begin{figure}[H]
\centering
\includegraphics[scale=0.6]{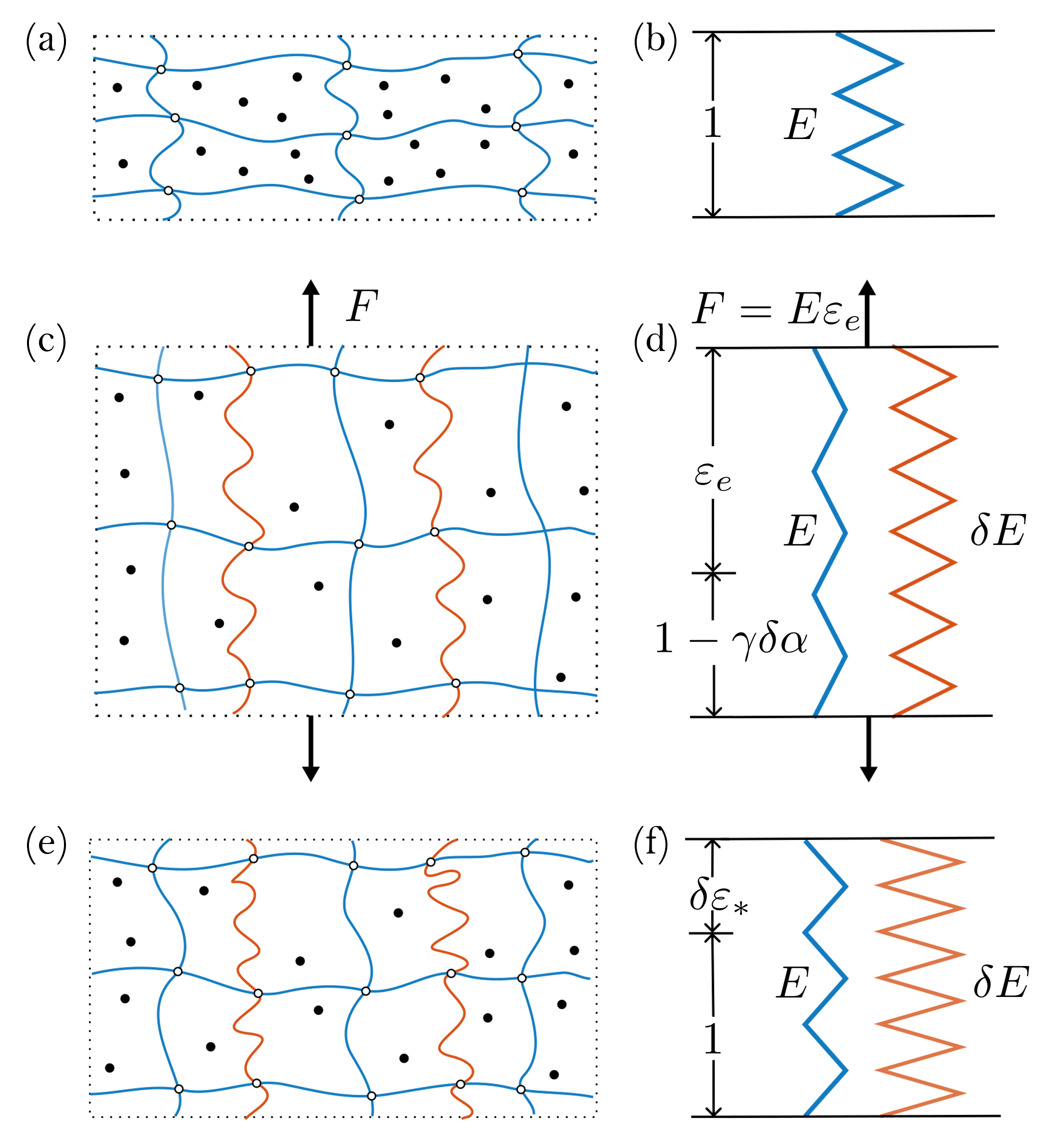}
\caption{Schematic illustration representing the mechanism governing evolution of the transformation strain. The polymer network is illustrated on the left (a,c,e) with equivalent spring system illustrated, respectively, on the right (b,d,f): (a) Initial polymer network composed of cross-linked polymer chains and free monomer molecules; (b) Equivalent initial spring system  with elastic modulus $E$ and unit stress-free length; (c) Polymerization under mechanical loading, red lines represents the newly formed stress-free polymer chains; (d) Mechanical loading extends the element, added parallel spring (red) is stress-free; (e) Upon unloading, the polymer network  reaches  a new stress-free state with internal force  balance between the old chains (blue) and the new chains (red); (f) The stress free system, represented by parallel springs has a new stress-free length.}
\label{fig:Transformation}
\end{figure}

As an illustrative example, consider a material element of unit initial length undergoing polymerization,  and  the equivalent spring system shown in Fig. \ref{fig:Transformation}. The initial stress-free state of the element and an equivalent  spring model with stiffness $E$ and stress-free length of $l=1$ are shown in  Fig. \ref{fig:Transformation}(a,b), respectively. Now, consider the element is subjected to a constant force, $F=E\varepsilon_e$, while the polymerization reaction leads to an incremental curing degree of $\delta \alpha$ in the stretched network (Fig. \ref{fig:Transformation}c,d), under isothermal conditions. To account for  chemical shrinkage caused by density difference between polymer and monomer, the stress-free length of the chains reduces by $\gamma\delta\alpha$, such that $l=1-\gamma\delta\alpha$, with $\gamma$ denoting the chemical shrinkage ratio. According to \eqref{modulus}, after the reaction, the stiffness of the element increases to $E+\delta E$, which is modelled by adding a parallel spring with incremental stiffness $\delta E$. 
An important assumption we make here is that the newly formed chains  of incremental stiffness $\delta E$, are added stress-free \citep{wu2018evolution}. Accordingly, their stress-free length is $l=1-\gamma\delta\alpha+\varepsilon_e$, while the pre-existing chains remain stretched by $\varepsilon_e$. 

Next, consider removing the applied load, as shown in Fig. \ref{fig:Transformation}(e,f). Since the two springs (or equivalently  polymer networks) have different stress-free lengths, the system will not return to the initial state ($l=1$). Instead, the two spring systems will mechanically  balance each other to find a new state that is macroscopically stress-free with unrecoverable mechanical energy, $\psi_*$, embedded into the material system. This incremental change of the stress-free state is described by   an increment of the transformation strain $\delta\varepsilon_*$ as shown in Fig. \ref{fig:Transformation}(f). The force balance between the two springs requires that
\begin{equation}
    E\cdot(\delta\varepsilon_*+\gamma\delta\alpha) + \delta E(\delta\varepsilon_*-\varepsilon_e+\gamma\delta\alpha) =0
\end{equation}
While the above result considers an incremental change in the polymerization process, we can neglect the higher order terms   and substitute \eqref{modulus} in to rewrite this equation in the differential form 
\begin{equation}\label{flow}
    \dot{\varepsilon}_* =  \left(\frac{E_H}{E_0+E_H\alpha}  \varepsilon_e-\gamma\right)\dot{\alpha}
\end{equation}
Note that under stress-free conditions ($\varepsilon_e = 0$), the above evolution law reduces to $\dot{\varepsilon}_* =-\gamma \dot\alpha$, which describes the chemical shrinking effect.

The above, purely mechanical, representation of the mechanism of transformation strain  excludes any potential mechanism of heat generation. Accordingly, we require that the forth term in \eqref{SL1} vanishes and thus that the inequality \eqref{heat_ineq} specializes to the relation 
\begin{equation}\label{network}
    \dot{\psi}_* = F \dot{\varepsilon}_*
\end{equation}
which describes the evolution of the network energy $\dot{\psi}_*$.

Note that in plasticity theory an analogous energy conversion relation defines the so-called plastic work $(\sigma\delta\varepsilon_p)$. However, the plastic work is considered to be not only transformed to the stored energy (typically called defect energy), but also converted into heat due to the dissipative nature of plastic deformation. The Taylor-Quinney coefficient is defined in plasticity theory as the fraction of the plastic work converted into heat, which is treated as a material constant. Accordingly, the relation in \eqref{network} is equivalent to prescribing  that the Taylor-Quinney coefficient in our system is  zero.

\subsection{Governing equations}
The theoretical model presented here can be described by the set of four independent  field variables $\{u,T,\alpha,\varepsilon_*\}(x,t)$ and four governing equations,
which after substitution of the specific constitutive relations take the final form:
 \begin{itemize}

 \item[\textit{I}.] {Mechanical equilibrium} (using \eqref{m_eq} with \eqref{const}):
\begin{equation}\label{govern_1}
    F_x =\left[(E_0+E_H\alpha)\varepsilon_e       \right]_x=0
\end{equation}
where we use \eqref{strain_decomposition} and \eqref{eT} to write the elastic strain as a function of these field variables, in the form 
\begin{equation}\label{e_e}
    \varepsilon_e = u_x-\beta T-\varepsilon_*
\end{equation}

 \item[\textit{II}.] Heat balance (using \eqref{heat0} with \eqref{const}  and \eqref{network}):
\begin{equation}\label{govern_2}
    c\dot T = \kappa T_{xx} -\lambda(T-T_0) + \mathcal F \dot\alpha -\beta T \dot{{F}}
\end{equation}
with  
\begin{equation}\label{reaction_heat_full}
    \mathcal{F} = H(1-\alpha) -\frac{1}{2}E_H \varepsilon_e^2 
\end{equation}
obtained from \eqref{const}.

 \item[\textit{III}.] Chemical kinetics:
\begin{equation}\label{govern_3}
    \dot\alpha = \hat{\mathcal{G}} (u,T,\alpha,\varepsilon_*)
\end{equation}
where $\hat{\mathcal{G}}$ is written in terms of the present field variables  by substituting the elastic strain \eqref{e_e} into \eqref{chem_kinetic}.

 \item[\textit{IV}.] Strain evolution (using   \eqref{const} and \eqref{flow}):
\begin{equation}\label{govern_4}
    \dot{\varepsilon}_* =  \left(\frac{E_H}{(E_0+E_H\alpha)^2}  F-\gamma\right)\dot{\alpha}
\end{equation}

\end{itemize}

Note that in absence of  mechanical effects (i.e. without \eqref{govern_1} and \eqref{govern_4}, such that $F\varepsilon_e \equiv 0$) our model would reduce to the typical reaction-diffusion system that has been previously considered by \cite{goli2018frontal}.

\subsection{Boundary and initial conditions}
To complete the representation of our system, it remains to define  boundary and initial conditions. 
Considering the system illustrated in Fig.\ref{fig:experiment}(a), we investigate the evolution of the  set of free field variables  in the interval $x\in [0,L_0]$ for $t\geq 0$. 

\vspace{3mm}
\noindent \textit{Boundary Conditions:} The uniform pre-stretch  exerted in the experiment is imposed by prescribing the displacement boundary conditions at the ends of the sample
\begin{equation}\label{b1}
    \quad  u(0,t) =0, \quad u(L_0,t) =\Delta
\end{equation}

To capture the thermal interaction between the sample and the grippers,  we assume the following Robin boundary condition  on the temperature 
\begin{equation}\label{b2}
    q(0,t) = -\Lambda(T(0,t)-T_0),\qquad  q(L_0,t) = \Lambda(T(L_0,t)-T_0)
\end{equation}
where $\Lambda$ is the heat transfer coefficient at the ends\footnote{Recall that flux is related to temperature gradients via Fourier's law \eqref{Fourier}.}.

\vspace{3mm}
\noindent \textit{Initial Conditions:} 
Before initiation of the FP process, we consider an un-reacted and pristine sample with 
\begin{equation}\label{init1}
   \alpha(x,0)  =0,\qquad \varepsilon_*(x,0) = 0
\end{equation}

To smoothly trigger the polymerization reaction while accounting for the initial heating of the soldering iron, the following initial condition is imposed on the temperature field 
\begin{equation}\label{init2}
    T(x,0) = (T_i-T_0) \exp\left(-\frac{x}{x_i}\right)+T_0
\end{equation}
where $T_i$ is the initiation temperature from  the soldering iron and $x_i$ represents the length scale of the heated region.

Finally, the equations \eqref{govern_1}, \eqref{govern_2}, \eqref{govern_3} and \eqref{govern_4}, together with the boundary conditions \eqref{b1}, \eqref{b2} and the initial conditions \eqref{init1}, \eqref{init2} completes the representation of our problem for the set of independent field variables  $\{u,T,\alpha,\varepsilon_*\}(x,t)$. 
 
\section{Numerical integration and  simulation results}
The above fully coupled uniaxial model has been solved numerically using a nonlinear finite element framework in Matlab. The time integration is performed using an implicit backward-Euler method. To accurately integrate across high spatial gradients in field variables at the polymerization front, a moving mesh method \citep{huang2010adaptive} has been applied. This method refines the mesh near the front at the end of each time step.


To reduce the complexity of the computation, while ensuring numerical accuracy, we employ a staggered integration strategy: At each time step, the reaction-diffusion equations \eqref{govern_2} and \eqref{govern_3} are first solved with the displacement and the transformation strain from the last time step. The mechanical equations \eqref{govern_1} and \eqref{govern_4} are then solved to update the displacement field and the transformation strain, and  are again used to  solve the reaction diffusion equations. This   process is repeated until convergence. For the FP system considered here, it can be shown that the   last term in the heat balance equation \eqref{govern_2} is negligible\footnote{Beyond basic scaling analysis, this has been validated by comparing the magnitudes of the separate contributions in  \eqref{govern_2}.  }; hence, we neglect this term in the simulations to further simplify the staggered integration process.

\begin{table}[h]
    \centering
    \begin{tabular}{llll}
    \hline
    Parameter & Value & Unit\\
    \hline
    $\tilde{c}$  & $1568$ &$\mathrm{kJ/(m^3\cdot K)}$ \\
        $\tilde{\kappa}$ &  $0.152$ & $\mathrm{W/(m\cdot K)}$ \\
     $\tilde{\lambda}$  & $16.2$ &$\mathrm{W/(m^2\cdot K)}$\\
     $\tilde{\Lambda}$  & $15.2$  &$\mathrm{W/(m^2\cdot K)}$\\
     $\tilde{H}$    & $3.67\times 10^5$ & $\mathrm{kJ/m^3}$\\
     $\beta$ & $10^{-4}$ & $\mathrm{K}^{-1}$\\
     $C$   & $1.8\times 10^{15}$   & $\mathrm{s}^{-1}$\\
     $E_a$ & 110.75  & $\mathrm{kJ/mol}$\\
     $\tilde{h}$  &  $2.20\times 10^5$   & $\mathrm{kJ/m^3}$\\
     $n$ & 1 & $-$\\
     $\tilde{E}_0$  & 200  & $\mathrm{kPa}$\\
     $\tilde{E}_H$  & 2  & $\mathrm{GPa}$\\
    \hline
    \end{tabular}
    \caption{Material parameters for FP of DCPD used in the simulations. All parameters are provided per unit volume, to exclude the dependence on the specific sample geometry, as denoted by the superimposed bar. Corresponding parameter values per unit length of the pre-streched sample are obtained by $ (~) =\tilde{(~)}\cdot A$, where  $A=0.5$ cm$^2$ is the deformed cross-sectional area of the cylindrical sample, used in all simulations. Note that the heat loss coefficient $\tilde{\lambda}$ is normalized by the lateral section area per unit length $\lambda =\tilde{\lambda}\cdot{\pi d}$, where $d$ is the diameter of the cross section }
    \label{tab:parameter}
\end{table}

To best represent the experimental system, the simulation is conducted in the entire region of the sample,  with initial length $L_0 =20\ \mathrm{cm}$. The material parameters used in the simulation are listed in Table \ref{tab:parameter} where each of the parameters is either obtained from the literature, or directly inferred from the experiments, as follows:

\begin{itemize}
  \item[--] The heat capacity $\tilde{c}$, the thermal conductivity $\tilde{\kappa}$, and the activation energy $E_a$, are adopted directly from \cite{lloyd2021spontaneous};
  
  \item[--] The chemical modulus $\tilde{H}$ is chosen to match the maximum temperature $T_{max}$ observed from the experiment;
  
  \item[--]  The heat loss coefficient $\tilde{\lambda}$ is determined experimentally (details in \ref{heat_loss}) ;
  
  \item[--] The tangent modulus of the gel $\tilde{E}_0$ is estimated from the relaxation experiments  shown in \ref{gel}, while the chemical hardening modulus $\tilde{E}_H$ and the thermal expansion coefficient $\beta$ are adopted from \cite{kumar2022surface}; 
  
  \item[--] The reaction rate constant $C$, the heat transfer coefficient $\tilde{\Lambda}$, the reaction order $n$, and the magnitude of the energy barrier $h$, are determined from the experimental observation shown in Fig. \ref{fig:experiment} (b). Note that each of these material parameters reflect different features in the response, thus allowing to infer them separately. For example, the propagation speed is influenced mainly by the value of the reaction rate constant $C$, while the heat transfer coefficient $\tilde{\Lambda}$ determines how fast the temperature decays at the boundary.
  

  
\end{itemize}

With the above material parameters defined, the only remaining  tuneable parameter is $\gamma$, which represents the chemical shrinkage ratio.  We will examine the influence of chemical shrinkage on the mechanical force response and the FP process in the next section, by considering various values of $\gamma$. It will be shown that for the present system shrinkage is negligible, i.e. $\gamma\sim 0$.
Since FP in absence of mechanical effects has been studied by \cite{goli2018frontal}, in what follows, we focus our simulation effort to elucidating the role of mechanical coupling. Specifically, we will examine the development of mechanical forces  during FP, and the influence of the stress on the propagation of the polymerization front. Through our analysis, we will establish that the numerical predictions of the fully coupled uniaxial model in Section 3 capture  the experimental observations described in Section 2.

\subsection{Mechanical force response}

\begin{figure} [H]
    \centering
    \includegraphics[width=1\columnwidth]{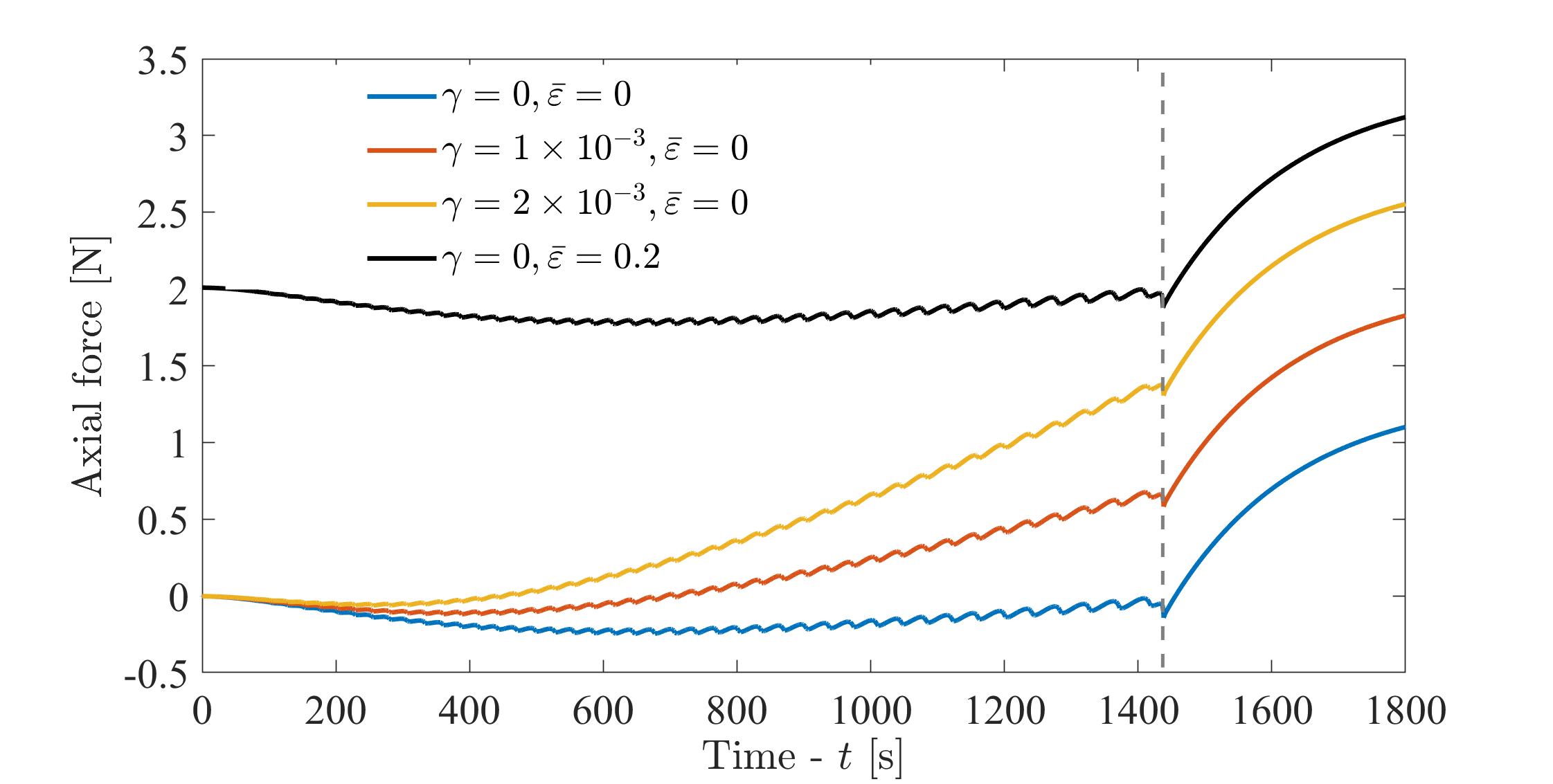}
    \caption{Simulation results: Axial force as a function of time for various values of the chemical shrinkage  ratio $\gamma$ and applied strain $\bar{\varepsilon}$.  The grey dashed
line indicates the time when the sample was fully polymerized.  }
    \label{fig:force_sim}
\end{figure}

The most obvious evidence of the mechanical coupling in FP is the change of axial force $F(t)$ during the polymerization process. The experimental force-time curve without applied strain ($\bar{\varepsilon}=0$) is presented in Fig.\ref{fig:experiment}(c). During front propagation, a compressive axial force develops until  a rapid transition occurs at the moment of full polymerization, followed by accumulation of  a tensile residual stress. 
This behavior is mirrored by the  simulation result in Fig. \ref{fig:force_sim} (blue curve). The remaining curves show the influence of   the  chemical shrinkage ratio $\gamma$ (red and yellow curves) and the applied strain $\bar{\varepsilon}$ (black curve). It is important to notice that in all cases, even after the sample has cooled back down to room temperature (i.e. $t=1800$ s), a tensile residual stress remains, thus indicating a change in the stress-free configuration of the sample, namely a transformation strain.     This transformation strain can emerge due to the chemical shrinkage or may be induced by stresses acting during polymerization, as shown in \eqref{flow}. 
By examining the curves for $\bar{\varepsilon}=0$ in Fig. \ref{fig:force_sim}, we find that  although chemical shrinkage induces residual stresses, it is not the only contributor, since even for the case with $\gamma=0$ the axial force eventually becomes positive.  We attribute this  to the internal stresses that are present during polymerization. Accordingly, the  compressive stress  induces a compressive  transformation strain that results in the overall contraction of the fully polymerized sample. Such development of residual stresses is expected to become more pronounced for larger samples and   can   thus have a profound  influence on the precision of fabrication and the performance of components in future industrial applications.

 Another important observation  obtained by comparing curves  for $\bar{\varepsilon}=0$ in Fig. \ref{fig:force_sim} with the experimental curve in Fig. \ref{fig:force_sim}
 is that with a non-zero $\gamma$, the axial force $F(t)$ becomes tensile (positive) during FP. This contrasts with the experimental observation where the force remains compressive. This discrepancy indicates that chemical shrinkage is negligible in this material system (i.e. $\gamma \ll 10^{-3}$) for FP in DCPD from a soft gel state. Consequently, we will choose $\gamma = 0$ for all following simulations.

 From Fig. \ref{fig:force_sim} we can also observe an oscillation of axial force during the FP process, which is directly related to the pulsation of the front temperature. Such oscillations of axial force do not appear in the experimental results shown in Fig. \ref{fig:experiment}(c). This discrepancy may arise from the 1D simplification of our  model. This underscores the necessity  to extend the current framework to 3D to capture more complex mechanical response.

In contrast to the shrinkage ratio, we find that the influence of applied strain on the  force response is simply a translation from the case with $\bar{\varepsilon}=0$, as seen from the curve for $\bar{\varepsilon}=0.2$  in Fig. \ref{fig:force_sim}. Note also  that all the curves in Fig. \ref{fig:force_sim} reach  full polymerization  at nearly the same time (grey dashed line). The non-intuitive alignment in polymerization time, despite the variations in force responses, implies a limited impact of mechanical forces on the velocity of propagation. This observation will be further examined in the next section, where we investigate the influence of mechanical force on the velocity of propagation.



\subsection{Stress-induced quenching of FP}
\begin{figure} [H]
    \centering
    \includegraphics[width=0.7\columnwidth]{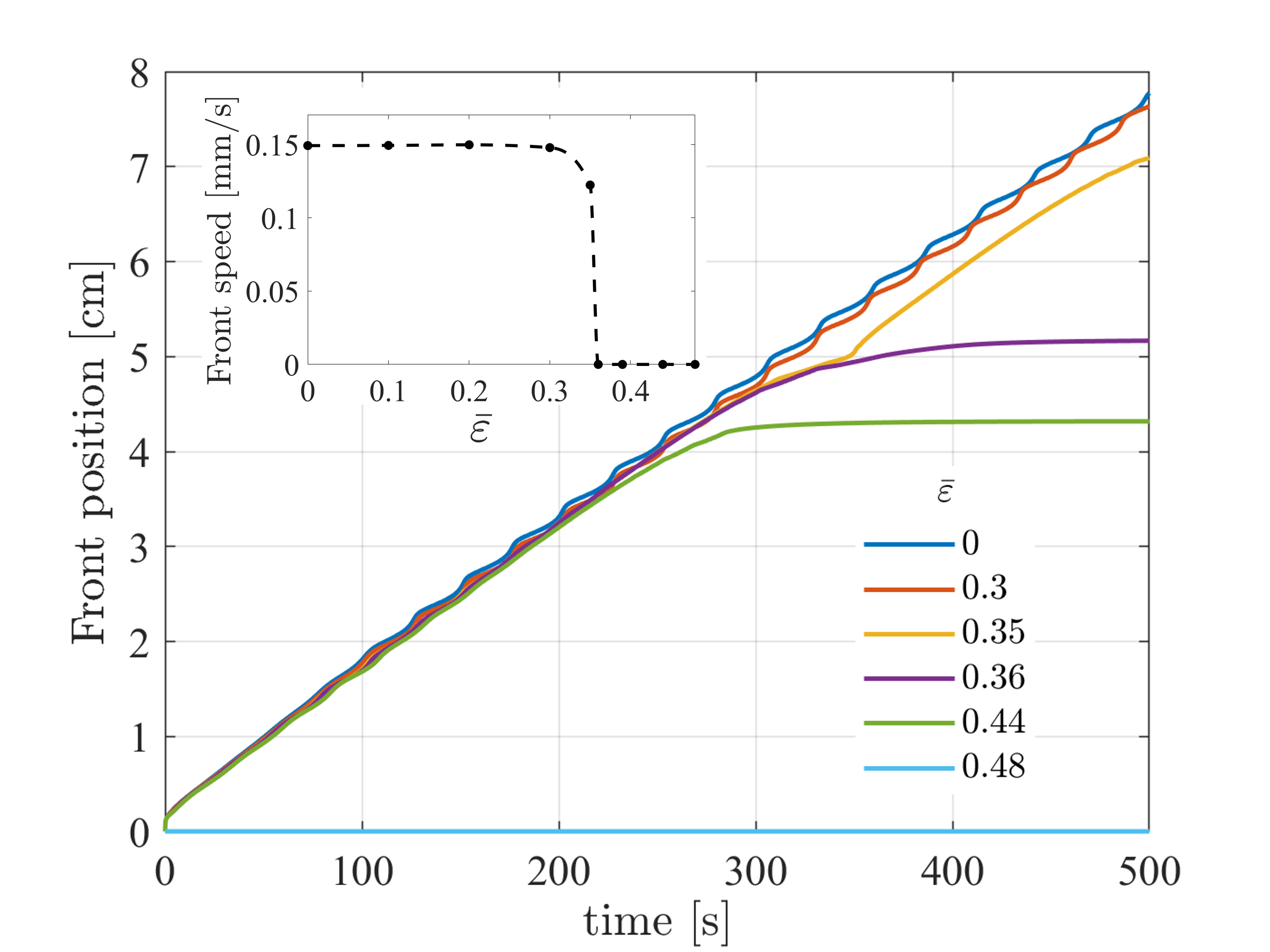}
    \caption{Simulation results: Front position as a function of time for various values of the applied strain $\bar{\varepsilon}$. Inset shows the propagation speed as a  function of the applied strain at $t=500 s$.}
    \label{fig:speed_sim}
\end{figure}

Next, we use our theoretical model to explain the experimental observation that mechanical forces can influence propagation dynamics and even quench FP. To this end, we conduct simulations for  various levels of the applied strain $\bar{\varepsilon}$ and  examine  the position of the front  as a function of time, as shown in Fig. \ref{fig:speed_sim}.


In absence of an applied strain ($\bar{\varepsilon}=0$), the front propagates at a nearly  constant  velocity, consistent with the experimental observation in Fig. \ref{fig:COMP}(a). Similar propagation speed is observed  for pre-strained samples up to a critical strain of  $\bar{\varepsilon}\sim 0.36$. Beyond this threshold,  the FP process initiates and propagates for some distance before ultimately quenching. This is consistent with the experimental result shown in Fig. \ref{fig:COMP}(b). While in the simulation the front propagates for $\sim 5 $ cm prior to quenching, in Fig. \ref{fig:COMP} the propagation quenches shortly after initiation. This discrepancy can be attributed to the difference of heating conditions between simulation and experiments: In the 1D framework the  cross section of the sample is heated uniformly to initiate the propagation, while in experiments only local heating can be achieved  through the point contact with the tip of a soldering iron.


To better illustrate the relation between the mechanical force and the propagation dynamics, the propagation speed at $t=500 $ s  is plotted as a function of the applied strain in the inset of Fig. \ref{fig:speed_sim}. We find a nearly binary  influence of the mechanical loading on the propagation dynamics:  Below the critical threshold the front propagates at a speed that is hardly influenced by the applied strain; beyond the critical threshold the propagation speed abruptly drops to zero.    We note that while this result, i.e. the critical value of $\bar{\varepsilon}\sim 0.36$, can be highly dependent on the chosen duration of the simulation, it nonetheless establishes that mechanical forces can quench FP, and that the longer the sample is the more it  is prone to this coupled effect. 

\begin{figure} [H]
    \centering
    \includegraphics[width=0.7\columnwidth]{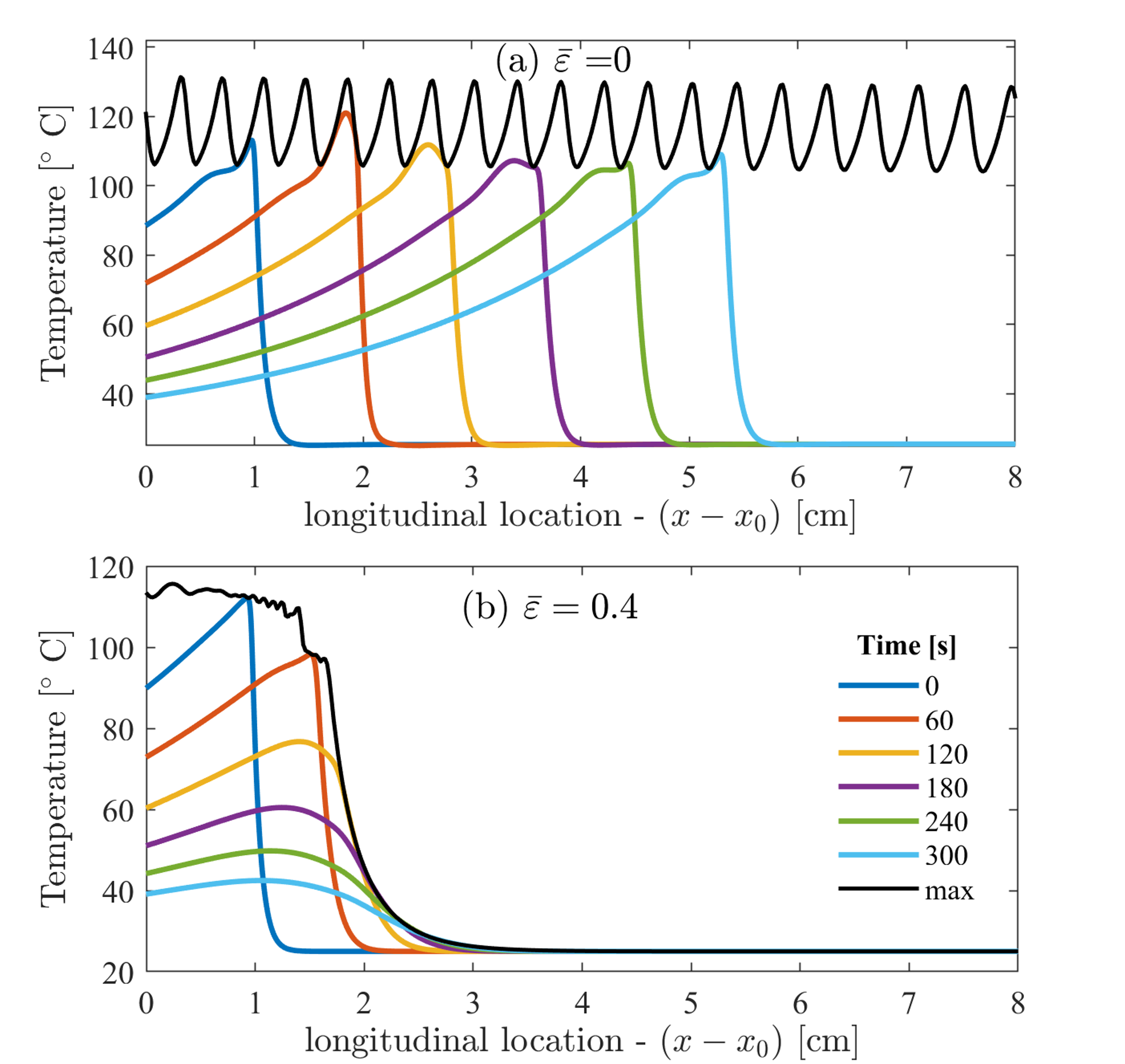}
    \caption{Simulation results: Evolution of temperature profile for the samples with (a): $\bar{\varepsilon}=0$; (b) $\bar{\varepsilon}=0.4$. Notice the shifted horizontal axis ($x-x_0$) with $x_0=5$[cm], which is chosen to best examine the region where the quenching occurs.} 
    \label{fig:comp_sim}
\end{figure}

 Next, to further illustrate how the propagation dynamics is influenced by the mechanical force, we show the evolution of temperature profiles with $\bar{\varepsilon}=0$ and $\bar{\varepsilon}=0.4$  in Fig. \ref{fig:comp_sim}. 
 The similarity  between the simulation result in Fig. \ref{fig:comp_sim} and the  experiment results in Fig. \ref{fig:COMP}, suggests  that our theoretical model  captures the experimentally observed  phenomenon.

While the above results confirm that the applied force influences the propagation dynamics and thus the chemical kinetics, the emergence of a nearly binary response, shown in Fig. \ref{fig:comp_sim}, is not obvious from the formulation of Section \ref{sec:theory}, as will be discussed next.

\subsection{Influence of mechanical force on chemical kinetics}

\begin{figure} [H]
    \centering
    \includegraphics[width=0.8\columnwidth]{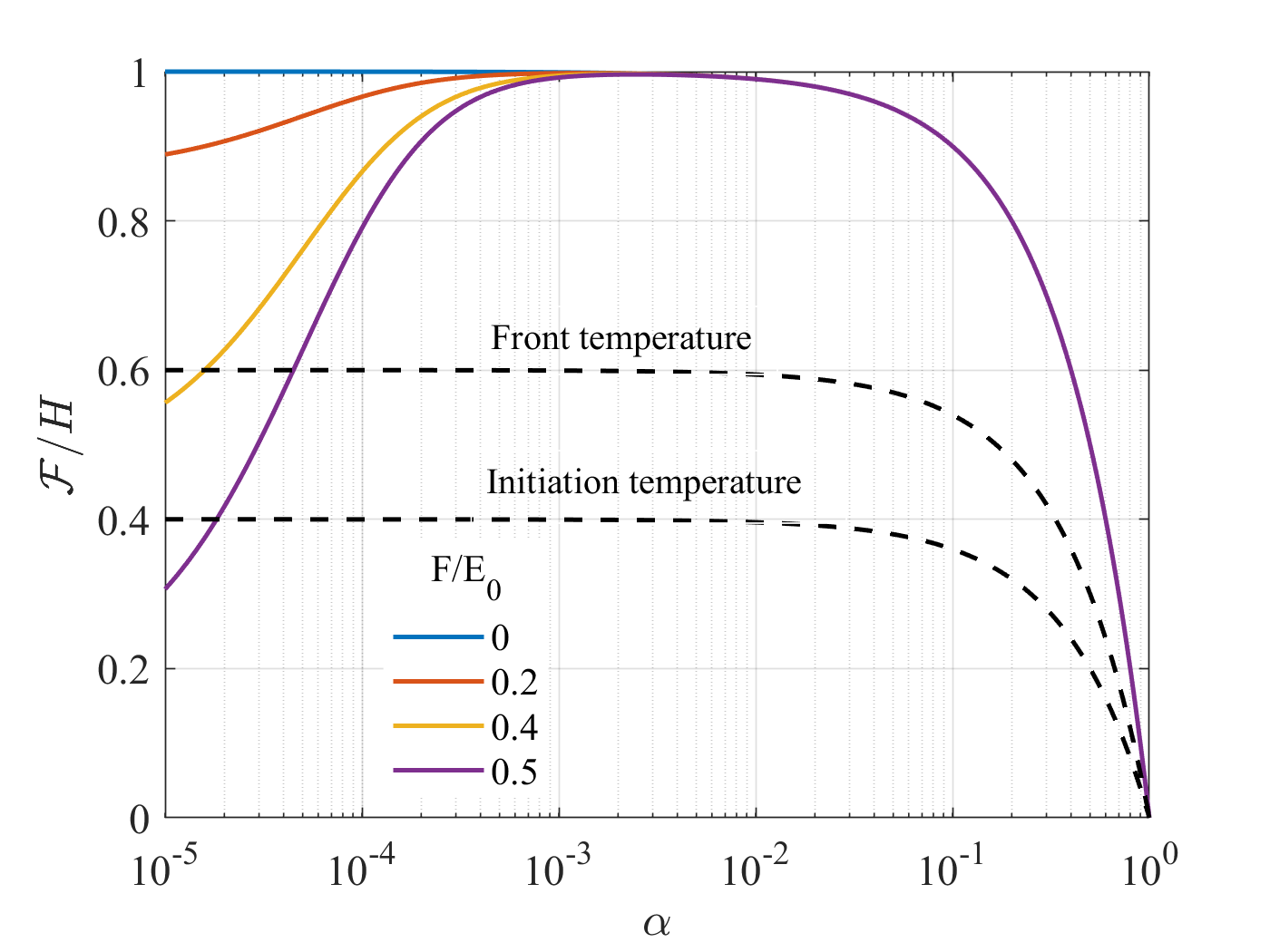}
    \caption{Driving force $\mathcal{F}$ as a function of the degree of curing $\alpha$ with various values of normalized axial force $F/E_0$. The black dashed line represents $\mathcal{F}_0$ defined in \eqref{barrier} calculated using the front temperature and the initiation temperature, as indicated. The degree of curing $\alpha$ is plotted on a logarithmic scale.}
    \label{fig:appendix_driving_force}
\end{figure}

To examine the relationship between  the driving force, $\mathcal{F}$, and the applied mechanical force, $F$, we begin by rewriting \eqref{reaction_heat_full} in terms of $F$ by substituting \eqref{const}$^1$ to arrive at the form
\begin{equation}\label{eq:driving}
    \mathcal{F}(\alpha, F) =H(1-\alpha) -  \frac{E_H}{2(E_0+E_H\alpha)^2 }F^2
\end{equation}
where the second term represents the mechanical coupling, which, as seen from the sign of the chemical hardening modulus $(E_H>0)$, will contribute to the reduction of  the driving force of FP, for any non-zero applied force. To further examine the consequences of this relation, we consider various values of constant applied force, and plot   the driving force  as a function of $\alpha$ in Fig. \ref{fig:appendix_driving_force}. 
Note that in absence of an applied mechanical force ($F=0$), the driving force is a linear function of $\alpha$. We find that the driving force is most significantly influenced by the applied force  at the initial stage of the reaction process ($\alpha \ll1$), and this influence becomes negligible shortly after, as $\alpha$ increases (note the logarithmic scale on the horizontal axis). 
Finally,  since the driving force is thermodynamically conjugate to the reaction rate ($\dot \alpha$), this implies that the reaction rate is insensitive to the applied force  for finite values of $\alpha$. This explains why, if not quenched, the propagation speed is insensitive to the applied force, as  observed in our simulation results (inset in Fig. \ref{fig:comp_sim}).  

  It remains to explain how the reduction of the driving force affects the change of propagation behavior and leads to quenching. According to the chemical kinetic relation \eqref{chem_kinetic}, the reaction can only occur when $\mathcal{F}\geq \mathcal{F}_0$, where $\mathcal{F}_0$ is a function of both  the degree of curing $\alpha$ and the temperature $T$, as defined in \eqref{barrier}. The FP process features two  values of temperature: the temperature used to initiate the propagation, $T_i$ - initiation temperature, and the characteristic temperature at the front, defined as\footnote{Here the value of front temperature is estimated by assuming that the reaction heat is fully transformed to local temperature increase.} $\hat{T}_f=T_0+H/c$. In Fig. \ref{fig:appendix_driving_force} we plot  $\mathcal{F}_0$  (dashed lines)  using both  the initiation temperature and the front temperature in  \eqref{barrier}.   
  As a result, we can now determine if the reaction can happen by comparing the values of $\mathcal{F}$ and $\mathcal{F}_0$ at the initial stage ($\alpha=0$), with a certain value of temperature. First, for the case without mechanical force ($F=0$), the driving force is always above the curves for  $\mathcal{F}_0$ with both initiation temperature and front temperature. This indicates that the FP process can not only be initiated but can also sustain itself in this case. Next, for the case with $F/E_0=0.4$, the value of $\mathcal{F}$ at $\alpha =0$ is larger than $\mathcal{F}_0$ with initiation temperature, but smaller than $\mathcal{F}_0$ with front temperature. This suggests that in this case the FP process can be initiated, but can not sustain itself, which is related to our experimental observation where the FP process initiates and propagates a finite distance before quenching. Lastly, for the case with $F/E_0=0.5$, the value of the driving force $\mathcal{F}$ at $\alpha=0$ is lower than $\mathcal{F}_0$ with initiation temperature, which means the FP process cannot be initiated. This observation is consistent with the simulation result plotted in Fig. \ref{fig:speed_sim}, and  explains  how the mechanical force can quench the FP process.


\section{Conclusions}
In the frontal polymerization process   a self-sustained  front propagates through the material driven by an exothermic reaction that  induces  sharp changes in the thermodynamic, chemical, and mechanical state of the  material. Existing models of FP either entirely neglect mechanical effects or assume that there is only a one-way coupling between mechanics, chemistry, and thermodynamics (i.e. that the propagation behavior of front is independent of the stress-state). In this work, we investigated the influence of the stress-state on the propagation behavior during FP combining  both  experiments and theory. By initiating frontal polymerization in stretched samples, our experimental results show that the propagation can be quenched by the application of mechanical force or by development of residual stresses in a sample, therefore suggesting that there is a two-way coupling effect between propagation dynamics and the  mechanical response during the FP process. To explain the experimental observation, We formulated a fully coupled thermo-chemo-mechanical theoretical model for FP in a uniaxial setting. Our model considers effects of  volume change  (i.e. thermal expansion and chemical shrinkage) and the change of mechanical stiffness during the FP process. A transformation strain is introduced in the kinematic representation to account for  evolution of the  stress-free configuration of the material akin to a plastic strain induced by the local stress-state upon polymerization. A driving force for the polymerization reaction is identified based on the second law of thermodynamics, from which a thermodynamically consistent chemical kinetic relation is prescribed. We implemented this model numerically and show that it is capable of explaining both the mechanical force response and the stress-induced quenching of FP from our experimental observations. 

A critical applied force emerges in both our experimental observations 
and  our theoretical model. Below this critical force propagation is observed at a nearly constant speed that is not affected by the applied force. Beyond this threshold, quenching is consistently observed. 
This binary response to  the mechanical force may explain why mechanical coupling in FP has not been previously appreciated. Nevertheless, when it comes to the fabrication of large-scale polymer composites (for example, air-craft wings), substantial residual stress will be accumulated during the FP process, which can not only induce significant undesired deformation in the fabricated components, but also quench the propagation. This highlights the piratical significance of understanding and controlling the mechanical coupling effects in FP for future large-scale industrial applications.

The ability of our model to account for the two-way coupling between the mechanical response and propagation dynamics offers dual utility: first, it can predict the potential residual stresses and deformation that are generated during the FP process, thereby facilitating design strategies to either avoid undesirable deformation or use the deformation to fabricate components with specific shape. Secondly, our work delves into the interplay between stress-states and propagation dynamics, providing a way to control the FP process through the application of mechanical force.
However, this work is not without limitations. Due to the absence of accurate measurement of several material parameters related to the chemical kinetics, the accuracy of our model still remains to be improved. Additionally, for applications involving large deformation, a finite elasticity model instead of the linear elastic model used currently is necessary to account for the nonlinear effects of mechanical response. Furthermore, the uniaxial experimental setup limits us to uniaxial stress-states, the potential effect of  more complex stress-states on the propagation dynamics still remains to be investigated, which requires experimental design beyond the current uniaxial setting. Therefore, future work lies in extending the current uniaxial linear elastic theoretical model and simulation to a 3D finite elastic framework. This will enable us to comprehensively explore the coupling phenomena associated with the FP process in scenarios involving both  complex geometries and nonlinear mechanical responses.

\section*{Acknowledgements} 
The authors would like to thank Professor John Pojman (LSU), and Dr. Jet Lem (MIT), for useful discussions. This work was partially supported by the Army Research Office, USA, under award no. W911NF-19-1-0275. 


\appendix
\section{Characterization of mechanical property of DCPD gel}
\label{gel}
\begin{figure} [h]
    \centering
    \includegraphics[width=1\columnwidth]{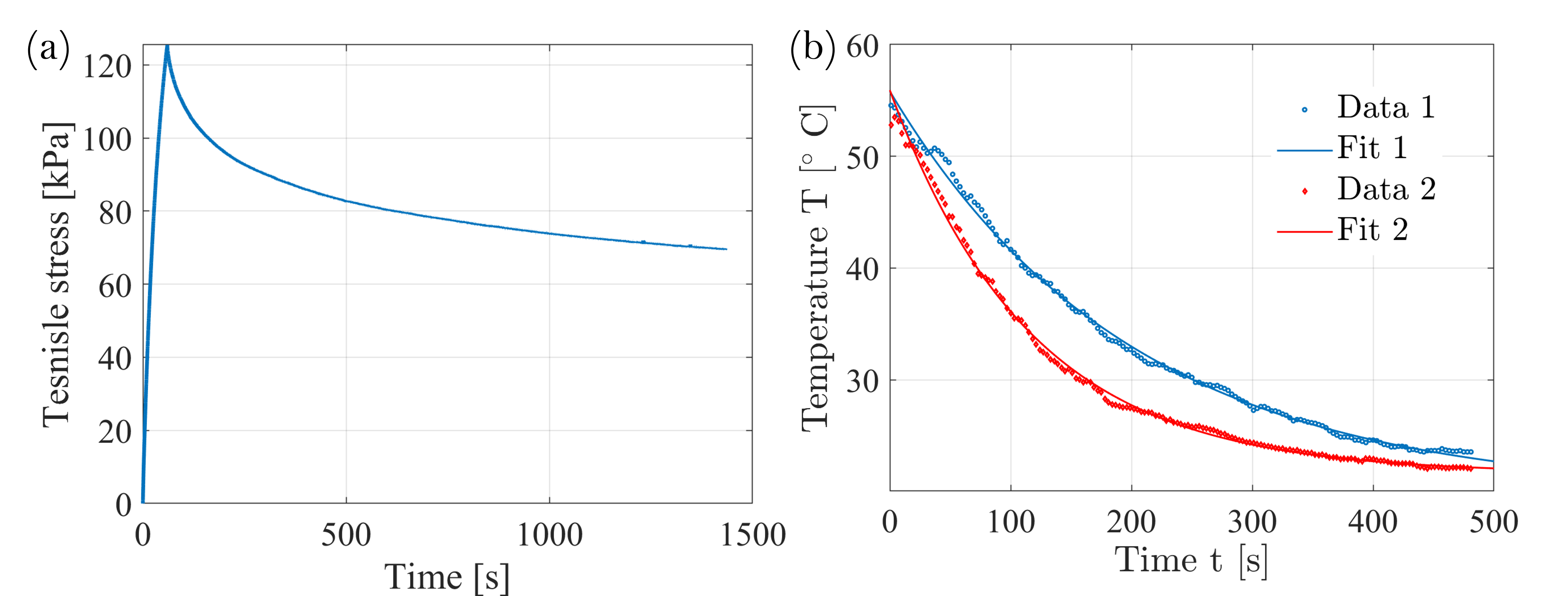}
    \caption{(a): Stress-time curve for DCPD gel in a relaxation test where the specimen was at first stretched to $\var{\varepsilon}=0.3$ with a constant strain rate $\dot{\varepsilon}=0.005\mathrm{s}^{-1}$ and then relaxed for half an hour; (b) Heat loss behavior (temperature $T$ as a function of time $t$) for two samples with different cross-section diameters: $7.4 \mathrm{mm}$ for sample 1 and $5.4 \mathrm{mm}$ for sample 2.}
    \label{fig:relaxation}
\end{figure}

To examine the visco-elastic behavior of the DCPD gel, here we performed a relaxation test on the specimen with initial cross-section diameter $8 \mathrm{mm}$. The sample was  first stretched to $\bar{\varepsilon} = 0.3$ with a constant strain rate $\dot{\varepsilon} = 0.005 \mathrm{s}^{-1}$ for the first sixty seconds, and then allowed to relax for  thirty minutes. The stress-time  curve is shown in Fig. \ref{fig:relaxation}(a).

The relaxation test reveals that the mechanical response of the DCPD gel shows rate dependence, which has not been taken into account in our current theoretical model. Form the value of the stress after relaxation we can also estimate the static modulus of the DCPD gel to be $\tilde{E}_0 =200\mathrm{kPa}$.

\section{Heat loss behavior}
\label{heat_loss}
To validate the assumption we made in \eqref{newton} that the heat loss behavior of our system follows the Newton's law of cooling and to measure the heat loss coefficient $\lambda$, we conducted an experiment to investigate the cooling behavior of the DCPD polymer (Note that here we assume that the thermodynamic properties don't change during the polymerization process).
Two short cylindrical polymer samples with different cross-section diameter ($7.4\mathrm{mm}$ for sample 1 and $5.4\mathrm{mm}$ for sample 2) were heated uniformly by water bath. The samples were then cooled under the same condition as the experiments shown in Fig.\ref{fig:experiment}(a), and the temperature change was recorded by the infrared camera.

Without any thermal diffusion and chemical reaction and under stress-free condition, the heat equation \eqref{govern_2} could be simplified as:
\begin{equation}\label{simplified_heat}
    c\dot{T} = -\lambda(T-T_0)
\end{equation}
which can be solved directly:
\begin{equation}
    T = (T_i-T_0) e^{-\frac{\lambda}{c} t}+T_0
\end{equation}
where $T_i$ is the initial temperature and $T_0$ is the room temperture.

Therefore, in Fig. \ref{fig:relaxation}(b) we presents the temperature  $T$  against time $t$ for the two samples together with the linear regression results. We can see a good match between the experimental data and the fitting results, which proves that the heat loss behavior follows the Newton's  law of cooling. 

According to the  regression result, the heat loss coefficient of sample 1 is $0.391 \mathrm{W/(m\cdot K)}$ while sample 2 is  $0.266 \mathrm{W/(m\cdot K)}$. Note that here the heat loss coefficient $\lambda$ is defined per unit length, which is dependent on the cross section area. To exclude such dependence on the geometry, here we introduce the heat transfer coefficient $\tilde\lambda$, which is related to the heat loss coefficient by $\lambda = \pi d \tilde{\lambda}$ with $d$ denotes the diameter of cross-section. For sample 1 we have $\tilde{\lambda}_1= 15.6\mathrm{W/(m^2\cdot K)}$, while for sample 2 we have $\tilde{\lambda}_1= 16.8 \mathrm{W/(m^2\cdot K)}$. As a result, we choose the averaged value $\tilde \lambda = 16.2\mathrm{W/(m^2\cdot K)}$ for the simulation.

\section*{Declaration of generative AI and AI-assisted technologies in the writing process}

During the preparation of this work the authors used ChatGPT in order to examine alternative language suggestions. After using this tool/service, the authors reviewed and edited the content as needed and take full responsibility for the content of the publication.





\bibliography{main}
\bibliographystyle{elsarticle-harv}
\end{document}